\documentclass[sigconf,numbers]{acmart}
\renewcommand\footnotetextcopyrightpermission[1]{} 
\AtBeginDocument{%
  }

\copyrightyear{2025}
\acmYear{2025}
\setcopyright{rightsretained}
\acmConference[WWW Companion '25]{Companion Proceedings of the ACM Web Conference 2025}{April 28-May 2, 2025}{Sydney, NSW, Australia}
\acmBooktitle{Companion Proceedings of the ACM Web Conference 2025 (WWW Companion '25), April 28-May 2, 2025, Sydney, NSW, Australia}
\acmDOI{10.1145/3701716.3715171}
\acmISBN{979-8-4007-1331-6/25/04}

\usepackage{blindtext}
\usepackage{xcolor}
\usepackage{amsmath}
\usepackage[utf8]{inputenc}
\usepackage[T1]{fontenc}
\usepackage{textcomp}
\usepackage{graphicx}
\usepackage{booktabs}
\usepackage{caption}
\usepackage{hyperref}

\begin{document}

\title{CyberAlly: Leveraging LLMs and Knowledge Graphs\\ to Empower Cyber Defenders}



\author{Minjune Kim}
\authornotemark[1]
\affiliation{%
  \institution{CSIRO's Data61}
  \city{Sydney}
  \state{NSW}
  \country{Australia}
}
\email{minjune.kim@data61.csiro.au}

\author{Jeff Wang}
\affiliation{%
  \institution{CSIRO's Data61}
  \city{Sydney}
  \state{NSW}
  \country{Australia}
}
\email{jeff.wang@data61.csiro.au}

\author{Kristen Moore}
\affiliation{%
  \institution{CSIRO's Data61}
  \city{Melbourne}
  \state{VIC}
  \country{Australia}
}
\email{kristen.moore@data61.csiro.au}

\author{Diksha Goel}
\affiliation{%
  \institution{CSIRO's Data61}
  \city{Melbourne}
  \state{VIC}
  \country{Australia}
}
\email{diksha.goel@data61.csiro.au}

\author{Derui Wang}
\affiliation{%
  \institution{CSIRO's Data61}
  \city{Melbourne}
  \state{VIC}
  \country{Australia}
}
\email{derek.wang@data61.csiro.au}

\author{Ahmad Mohsin}
\affiliation{%
  \institution{Edith Cowan University}
  \city{Perth}
  \state{WA}
  \country{Australia}
}
\email{a.mohsin@ecu.edu.au}

\author{Ahmed Ibrahim}
\affiliation{%
  \institution{Edith Cowan University}
  \city{Perth}
  \state{WA}
  \country{Australia}
}
\email{ahmed.ibrahim@ecu.edu.au}

\author{Robin Doss}
\affiliation{%
  \institution{Deakin University}
  \city{Melbourne}
  \state{VIC}
  \country{Australia}
}
\email{robin.doss@deakin.edu.au}

\author{Seyit Camtepe}
\affiliation{%
  \institution{CSIRO's Data61}
  \city{Sydney}
  \state{NSW}
  \country{Australia}
}
\email{seyit.camtepe@data61.csiro.au}

\author{Helge Janicke}
\affiliation{%
  \institution{Edith Cowan University}
  \city{Perth}
  \state{WA}
  \country{Australia}
}
\email{h.janicke@ecu.edu.au}

\renewcommand{\shortauthors}{Minjune Kim et al.}

\begin{abstract}
The increasing frequency and sophistication of cyberattacks demand innovative approaches to strengthen defense capabilities. Training on live infrastructure poses significant risks to organizations, making secure, isolated cyber ranges an essential tool for conducting Red vs. Blue Team training events. These events enable security teams to refine their skills without impacting operational environments. While such training provides a strong foundation, the ever-evolving nature of cyber threats necessitates additional support for effective defense. To address this challenge, we introduce CyberAlly, a knowledge graph-enhanced AI assistant designed to enhance the efficiency and effectiveness of Blue Teams during incident response. Integrated into our cyber range alongside an open-source SIEM platform, CyberAlly monitors alerts, tracks Blue Team actions, and suggests tailored mitigation recommendations based on insights from prior Red vs. Blue Team exercises. This demonstration highlights the feasibility and impact of CyberAlly in augmenting incident response and equipping defenders to tackle evolving threats with greater precision and confidence.

\end{abstract}




\maketitle

\section{Introduction}

\let\svthefootnote\thefootnote
\let\thefootnote\relax
\footnotetext{The manuscript has been accepted by WWW Companion ’25 Demo Track}
\let\thefootnote\svthefootnote
The increasing complexity and frequency of cyberattacks present significant challenges for cybersecurity practitioners. Modern Security Operation Centers (SOCs) must process vast volumes of system logs and alerts, the majority of which are benign or redundant -- false positives arising from benign events being misinterpreted as potential attacks. This pervasive issue diminishes the efficiency of cyber defenders, leading to ‘alert fatigue’ and increasing the risk of critical threats going undetected. While existing Machine Learning (ML) approaches have shown promise in filtering and classifying alerts, they often lack the contextual awareness necessary to generate actionable insights. Moreover, effectively integrating human expertise with automated systems remains a critical gap in current cybersecurity solutions.

To address these challenges, we focused on the maritime sector in developing the cyber range for our training exercises, recognising the rise in attacks targeting this critical infrastructure~\cite{senarak2024port}. The range replicates the IT and operational technology (OT) systems of a shipping port, providing a realistic and secure environment for conducting Red vs. Blue Team wargames, where the Red Team simulates attackers and the Blue Team defends against threats. By simulating real-world attack scenarios, the cyber range supports Blue Teams in refining their skills while exploring innovative solutions to bridge the gap between human expertise and automated systems. Our demo addresses key research and engineering challenges to aid Blue Team members in combating cyber threats with an effective human-AI collaboration. These challenges are outlined as follows:

\textbf{Challenge 1: Minimising Duplicated and Benign Alerts?} Enterprise networks and OT systems in the cyber range generate a substantial volume of logs and alerts, much of which constitutes background noise. This noise often causes confusion, making it difficult to distinguish legitimate activity from actual attacks. Prior studies~\cite{VAARANDI_CSR2021, SHITTU_LCN2014} have shown that a high proportion of daily alerts in enterprise networks are duplicates, with only a small subset representing suspicious activity. ML-based methodologies have been explored to address this issue by reducing duplication or classifying alerts. For example, Kidmose et al. 
 ~\cite{Kidmose_IEEEAccess2020} proposed a supervised learning method using recurrent neural networks (RNNs) and semantic analysis to map textual alerts to vector representations. This mapping allowed the identification of incidents by clustering similar alerts in vector space. Other works have employed correlation~\cite{Mirheidari_Note2013} and alert clustering~\cite{Lin_KDD2014, Zhao_ICSE2020} to reduce insignificant alerts. 
 Classification methods have also been proposed to identify attacks based on textual similarity~\cite{KIM_QRS2022, VAARANDI_CNSM2010}. 

\textbf{Challenge 2: Generating Intelligent Suggestions with Contextual Awareness?} Incident alerts often lack the contextual information necessary for security analysts to fully understand and address threats. Intelligent systems can enhance defenders' decision-making by integrating alerts with system-based contextual data, such as the history of related incidents or predefined playbooks. This additional context helps ensure that suggestions are both relevant and actionable in real-time scenarios. Modern large language models (LLMs) offer advantages for solving complex problems using general system knowledge. However, addressing specialised downstream tasks in the cybersecurity domain requires additional techniques, such as fine-tuning and prompt engineering. Retrieval-Augmented Generation (RAG)~\cite{lewis2020retrieval} provides a framework to enhance LLMs by retrieving relevant knowledge from external sources and incorporating it into the generation process. This approach combines domain-specific knowledge with real-time contextual information to improve the quality and relevance of outputs~\cite{webb2024cyber}. In our implementation, we utilise an LLM with a knowledge-graph RAG (KG-RAG) approach to combine general system knowledge with multiple knowledge sources and real-time data, enabling more precise and contextually aware suggestions.

\textbf{Challenge 3: Ensuring Comprehensive Validation of Proposed Methods?} Evaluating the outputs generated by LLMs in the cybersecurity domain presents significant challenges due to the limited availability of relevant datasets. Existing resources, such as CyberMetric~\cite{Tihanyi_CyberMetric2024}, which provides AI-driven questions and answers, and CyberBench~\cite{Liu_CyberBench2024}, a multi-task benchmark for tasks like recognition and summarization, offer valuable starting points for evaluation. Similarly, Bhusal et al.~\cite{Bhusal_ACSAC2024} introduced a benchmark designed to assess LLM performance in realistic cybersecurity advisory contexts. However, these datasets do not fully capture the complexities and nuances of our specific use cases. To address this limitation, we conducted data collection during our Red vs. Blue Team events. This process included gathering system logs, expert responses, and alert messages, enabling us to construct a ground truth dataset tailored to our system-specific environments and scenarios. This dataset serves as a robust foundation for evaluating our methods within the practical context of cybersecurity operations.
\vspace{-3mm}

\begin{figure}[h!]
  \includegraphics[width=0.4\textwidth]{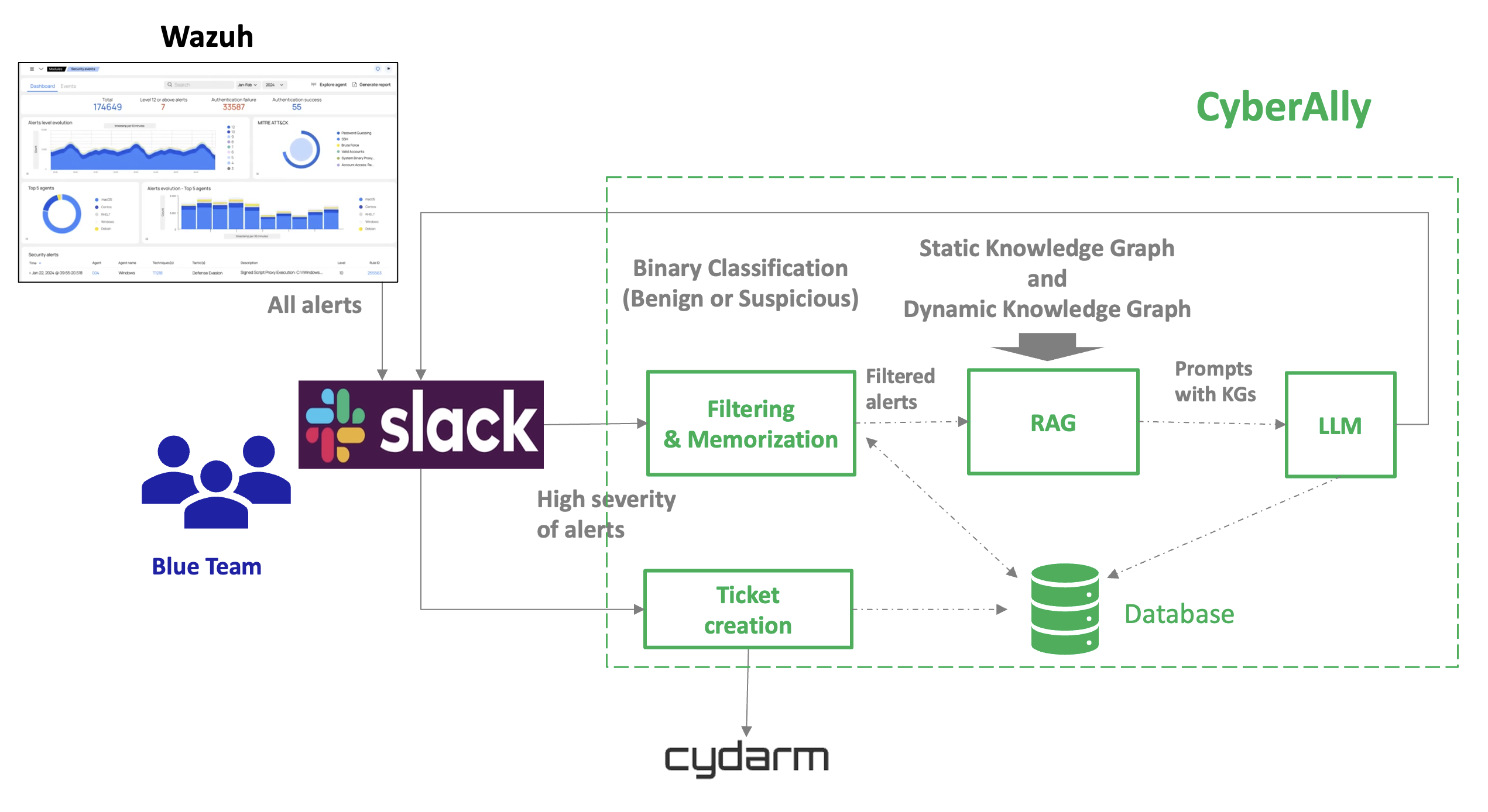}
  \caption{\footnotesize{CyberAlly Workflow: The system automatically analyses a high volume of incoming alert messages for the Wazuh SIEM\protect\footnotemark[1] via Slack\protect\footnotemark[2], generating intelligent suggestions for the Blue Team based on static and dynamic graph knowledge, and automates case ticket creation in Cydarm\protect\footnotemark[3].}}
  \label{fig:worflow}
\vspace{-5mm}
\end{figure}
\footnotetext[1]{https://wazuh.com/}
\footnotetext[2]{https://slack.com/}
\footnotetext[3]{https://www.cydarm.com/}

\section{Methodologies and System Design}

This paper introduces CyberAlly, an LLM-driven AI assistant that integrates knowledge graphs and LLMs to enhance incident response capabilities. During one of our 2-day training events, we collected a dataset of more than 60 thousand alert messages for training and verification. Of these, around 2.5 thousand distinct alert types were detected, excluding background noise, as summarized in Table~\ref{tab:alerts}. The table highlights the importance of removing duplicates and noise to enable effective response. The system architecture consists of two major components: alert filtering and dynamic prompting with knowledge graphs, as illustrated in Figure~\ref{fig:worflow}. For alert filtering, we utilize a k-nearest neighbours (kNN) model for binary classification. For dynamic prompting, we implement a KG-RAG using LlamaIndex, enabling contextually aware responses.  Additionally, we leverage the Slack Bolt library~\footnote[4]{https://api.slack.com/bolt} to ensure seamless user interactions with the system.

\begin{table}[h!]
\tiny
\resizebox{0.48\textwidth}{!}{%
\begin{tabular}{|c|c|c|c|c|c|c|c|c|c|c|}
\hline
Priority & 3 & 4 & 5 & 6 & 7 & 8 & 9 & 10 & 12 & 15\\
\hline
 \# incl. dup & 190 & 11 & \textbf{1461} & 20 & 109 & 5 & 47 & 612  & 1 & 55\\
 \# excl. dup & \textbf{82} & 6 & 78 & 6 & 15 & 1 & 4 & 16 & 1 & 1\\
\hline
\end{tabular}
}
\caption{\footnotesize{\#Alert messages: The first row shows the total number of alerts, including duplicates, while the second row indicates the count after duplicates are removed.}}
\label{tab:alerts}
\vspace{-10mm}
\end{table}

\subsection{Alert Filtering and Classification}
Managing alert overhead is a critical challenge for cyber defenders, who often struggle to distinguish between benign and malicious activity amidst numerous alert messages. The simulated environment in the cyber range, designed to replicate real-world networks, inherently generates false alerts from background activity. To address this issue, our system employs a two-step word-embedding-based approach: filtering duplicate alerts and categorising them using binary classification model.

\textbf{Duplicate Alert Filtering.}
To reduce alert volume, the system leverages sentence embeddings to map alert messages into a latent vector space. Input text for vectorization includes both full log details and titles, with embeddings calculated as the average of word vectors. 
Cosine distance is then measured, and duplicates are excluded if their similarity score exceeds a predefined threshold. This filtering process reduces redundancy and computational overhead.

\begin{table}[h]
\resizebox{0.48\textwidth}{!}{%
\begin{tabular}{|c|c|c|c|c|}
\hline
 &Benign only&Benign + 1x Malicious&Benign + 5x Malicious&Benign + 10x Malicious\\
\hline
 Precision & 0.9392 & 0.9904 & 0.9981 & 0.9992 \\
 Recall & 0.9679 & 0.9974 & 0.9995 & 0.9997\\
 F1 & 0.9533 & 0.9939 & 0.9988 & 0.9994\\
\hline
\end{tabular}
}
\caption{\footnotesize{Performance of the binary classification model with rebalanced datasets by upsampling malicious data instances at scales of 1x, 5x and 10x.}}
\label{tab:accuracy}
\vspace{-5mm}
\end{table}

\textbf{Binary Classification with Feature Extraction.} After filtering, a binary classification model categorizes the remaining alerts into two classes: benign or suspicious. As in the de-duplication process, semantic features are extracted from alert messages through the sentence embedding process, and the resulting embedding vectors are fed into the model, ensuring it retains critical contextual information for effective classification. The classification process employs a k-nearest neighbors (kNN) model with k=15, a value that demonstrated the best performance on our dataset. To address class imbalance, malicious samples were upsampled using a weighting technique, ensuring equitable representation during training. Additionally, a time window of 30 minutes was configured to capture temporal patterns in the alert data. The model's accuracy was evaluated using 10-fold cross-validation to mitigate overfitting, with results summarized in Table~\ref{tab:accuracy}.  Alerts classified as benign are discarded, while those deemed suspicious progress to the next stage, where they are analyzed by the KG-RAG LLM system. This integration enables defenders to focus on critical threats, minimizing false positives and enhancing response efficiency.


\subsection{Dynamic Prompting using KG-RAG}
\begin{figure}[hbt!]
  \includegraphics[width=0.35\textwidth]{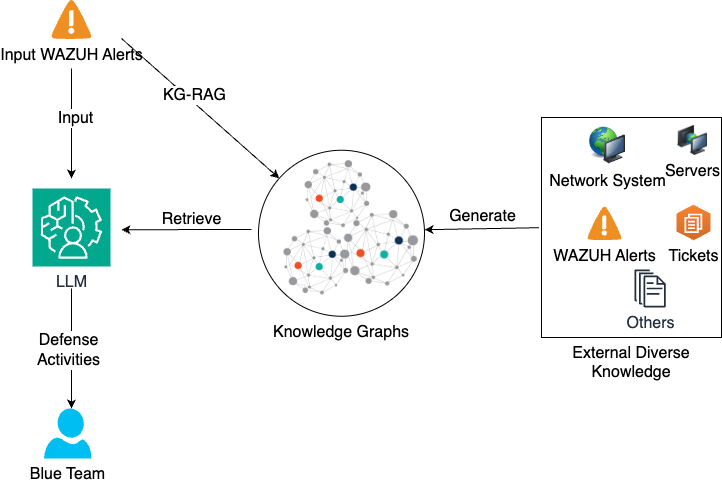}
  \caption{\footnotesize{LLM with KG-RAG - Knowledge Graphs capture the relationships between past and present events and enhance contextual understanding.
}}
  \label{fig:rag}
  \vspace{-3mm}
\end{figure}

Despite advancements in LLMs, they often face challenges with domain-specific problems due to limited specialized knowledge and contextual understanding. To address these limitations, we propose a Knowledge Graph-Enhanced Retrieval-Augmented Generation (KG-RAG) framework, as shown in Figure~\ref{fig:rag}, which integrates knowledge graphs to retrieve more accurate and context-specific data. The proposed knowledge graph architecture consists of two sub-graphs: a static graph representing knowledge from previous Red vs. Blue Team events, and a dynamic graph representing the current Red vs. Blue Team event and its alerts and Blue Team case management and change management tickets. Both the static and dynamic graphs contain the cyber range infrastructure and architecture details. With these sub-graphs, the KG-RAG can leverage both historical data and the current system state to inform the LLM's generation process.  This integration captures nuanced relationships between system components, enhancing the contextual accuracy of outputs.

We implement the AI assistant using GPT-4o developed by OpenAI\footnote[5]{https://openai.com/} and KG-RAG to feed relevant information from internal and external sources (e.g., MITRE\footnote[6]{https://attack.mitre.org/} frameworks). This enables the system to stay up-to-date with current events and domain-specific contexts, improving the precision and relevance of its responses. KG-RAG enhances LLM-generated suggestions by dynamically enriching the input prompt with relevant contextual information. To retrieve the most relevant information, KG-RAG computes similarity scores between the filtered incoming SIEM alert and the static and dynamic graphs, utilising LlamaIndex\footnote[7]{https://www.llamaindex.ai/}. 

Our architecture extends standard knowledge graph-based methods by incorporating dynamic updates to reflect real-time changes in the system. This approach allows the LLM to adapt to evolving conditions and provide more accurate and contextually relevant suggestions. Our preliminary experiments suggest that our KG-RAG module holds promise for addressing domain-specific tasks by leveraging both static and dynamic graphs. This approach enables the creation of a robust and adaptable system designed to enhance the accuracy and relevance of domain-specific outputs in real-time cybersecurity contexts.


\begin{figure*}[hbt!]
  \includegraphics[width=0.8\textwidth]{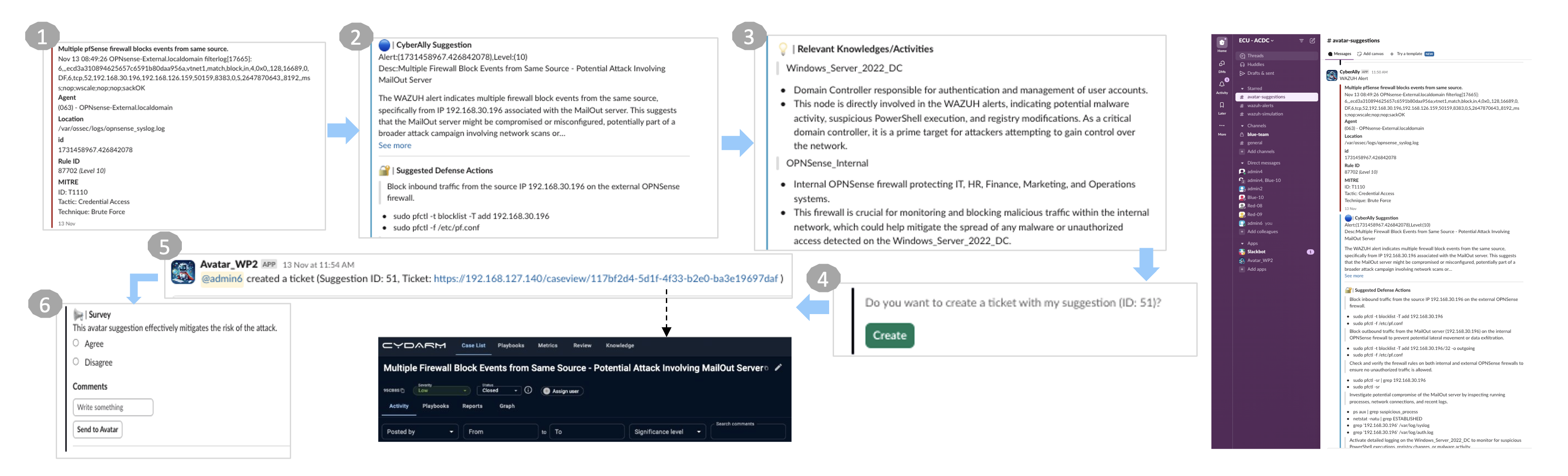}
  \caption{\footnotesize{CyberAlly in Slack. CyberAlly demo follows these steps: 1) monitor Wazuh alerts, 2) generate descriptions and suggest actions, 3) provide reasoning, 4) support decision-making, 5) create tickets in the Cydarm case management system, and 6) collect feedback.}}
  \label{fig:workflow}
  \vspace{-5mm}
\end{figure*}

\subsection{Slack Integration}
The Cyber Security Incident Response Plan\footnote[8]{https://www.cyber.gov.au/resources-business-and-government/governance-and-user-education/incident-response/cyber-security-incident-response-planning-practitioner-guidance} emphasizes the importance of effective logistics and communication tools, such as Slack, in coordinating incident response. During Red vs. Blue Team training events, the Blue Team relies primarily on Slack channels for communication. To enhance this workflow, CyberAlly has been integrated with Slack using a websocket, ensuring seamless and consistent interaction. As illustrated in Figure~\ref{fig:workflow}, we utilise Slack's block builder to create customized message blocks that provide a comprehensive overview of incoming alerts. These blocks comprise three main components: \\
\textbf{i) Alert summary:} Offers contextual insight into the alert, including any relevant connections to prior alerts or similar incidents.,\\
\textbf{ii) Recommended Actions: }Provides actionable insights and mitigations, including command-line recommendations to address potential threats.\\
\textbf{iii) Explanation and Reasoning:} Delivers transparent reasoning for the suggested actions, empowering users to make informed decisions. 
\vspace{-4mm}
\section{Demonstration}
 Figure~\ref{fig:workflow} illustrates the key user interactions in the demonstration. During the real-time demo, participants will take on the role of Blue Team members and interact with CyberAlly to respond to simulated cybersecurity incidents. Alerts captured during previous events will be replayed to simluate a live environment, enabling participants to experience the functionality of CyberAlly in action. The primary workflows driving CyberAlly's suggestions are as follows:

\begin{enumerate}
    \item \textbf{Wazuh Alert Monitoring:} Participants will observe Wazuh posting alerts to Slack, where CyberAlly analyzes incoming data for suspicious behaviors. The system automatically filters out alerts identified as duplicates or benign based on similarity thresholds.
    \item \textbf{Description \& Suggested Actions:} Participants will receive actionable recommendations from CyberAlly to address ongoing attacks. By leveraging static and dynamic knowledge graphs, the system generates defence suggestions, including shell script-based commands, to guide participants in mitigating threats.
    \item \textbf{Reasoning:} CyberAlly will provide participants with explanations behind its recommendations, offering insights into the underlying reasons and relationships that inform its suggestions. This includes references to network architecture and historical events.
    \item \textbf{Decision Making:} Participants will decide whether to create task tickets for handling incidents. CyberAlly facilitates this process by automatically populating tickets with recommended actions and relevant contextual information.
    \item \textbf{Cydarm Ticket Creation:} Participants will see how CyberAlly generates and links task tickets in the Cydarm case management system, enabling efficient handling of suspicious alerts.
    \item \textbf{Feedback Collection:} At the end of the demo, participants will provide feedback on CyberAlly's performance, which will be collected to inform ongoing research and future improvements.
\end{enumerate}
\vspace{-2mm}
This hands-on experience allows participants to explore how CyberAlly enhances Blue Team workflows, demonstrating its ability to streamline decision-making and improve incident response effectively.

\vspace{-4mm}
\section{Ethics}
This research adhered to ethical guidelines, with approval obtained from the ethics review boards of participating institutions. Participants in the cybersecurity wargame events provided informed consent prior to participation. During the demonstration at the conference, no data will be collected, but participants will receive a consent form detailing the nature of their involvement to ensure transparency and alignment with ethical best practices.

\vspace{-2mm}
\section{Conclusion}
This paper introduces a system designed to enhance human-AI collaboration within a cybersecurity wargame setting. CyberAlly is an intelligent suggestion system that assists cyber defenders by analyzing incoming alerts and providing context-aware recommendations for effective incident response. Integrating CyberAlly into a practical cyber range demonstrates its potential to reduce workloads in SOCs while offering actionable insights linked to prior or related activities. This system represents a step forward in leveraging AI to support and augment human decision-making in complex cybersecurity environments.

\vspace{-1mm}
\begin{acks}
The work has been supported by the Cyber Security Research Centre Limited, whose activities are partially funded by the Australian Government’s Cooperative Research Centres Programme.
\end{acks}
\vspace{-1mm}
\bibliographystyle{ACM-Reference-Format}
\tiny
\bibliography{ref.bib}


\begin{thebibliography}{14}


\ifx \showCODEN    \undefined \def \showCODEN     #1{\unskip}     \fi
\ifx \showDOI      \undefined \def \showDOI       #1{#1}\fi
\ifx \showISBNx    \undefined \def \showISBNx     #1{\unskip}     \fi
\ifx \showISBNxiii \undefined \def \showISBNxiii  #1{\unskip}     \fi
\ifx \showISSN     \undefined \def \showISSN      #1{\unskip}     \fi
\ifx \showLCCN     \undefined \def \showLCCN      #1{\unskip}     \fi
\ifx \shownote     \undefined \def \shownote      #1{#1}          \fi
\ifx \showarticletitle \undefined \def \showarticletitle #1{#1}   \fi
\ifx \showURL      \undefined \def \showURL       {\relax}        \fi
\providecommand\bibfield[2]{#2}
\providecommand\bibinfo[2]{#2}
\providecommand\natexlab[1]{#1}
\providecommand\showeprint[2][]{arXiv:#2}

\bibitem[Bhusal et~al\mbox{.}(2024)]%
        {Bhusal_ACSAC2024}
\bibfield{author}{\bibinfo{person}{Dipkamal Bhusal}, \bibinfo{person}{Md~Tanvirul Alam}, \bibinfo{person}{Le Nguyen}, \bibinfo{person}{Ashim Mahara}, \bibinfo{person}{Zachary Lightcap}, \bibinfo{person}{Rodney Frazier}, \bibinfo{person}{Romy Fieblinger}, \bibinfo{person}{Grace~Long Torales}, {and} \bibinfo{person}{Nidhi Rastogi}.} \bibinfo{year}{2024}\natexlab{}.
\newblock \showarticletitle{SECURE: Benchmarking Generative Large Language Models for Cybersecurity Advisory}.
\newblock \bibinfo{journal}{\emph{arXiv preprint arXiv:2405.20441}} (\bibinfo{year}{2024}).
\newblock


\bibitem[et~al.(2020)]%
        {Zhao_ICSE2020}
\bibfield{author}{\bibinfo{person}{Nengwen et al.}} \bibinfo{year}{2020}\natexlab{}.
\newblock \showarticletitle{{Understanding and handling alert storm for online service systems}}.
\newblock \bibinfo{journal}{\emph{Proceedings - International Conference on Software Engineering}} (\bibinfo{year}{2020}).
\newblock
\showISBNx{9781450371230}
\showISSN{02705257}


\bibitem[Kidmose et~al\mbox{.}(2020)]%
        {Kidmose_IEEEAccess2020}
\bibfield{author}{\bibinfo{person}{Egon Kidmose}, \bibinfo{person}{Matija Stevanovic}, \bibinfo{person}{Søren Brandbyge}, {and} \bibinfo{person}{Jens~M. Pedersen}.} \bibinfo{year}{2020}\natexlab{}.
\newblock \showarticletitle{Featureless Discovery of Correlated and False Intrusion Alerts}.
\newblock \bibinfo{journal}{\emph{IEEE Access}}  \bibinfo{volume}{8} (\bibinfo{year}{2020}), \bibinfo{pages}{108748--108765}.
\newblock


\bibitem[Kim et~al\mbox{.}(2022)]%
        {KIM_QRS2022}
\bibfield{author}{\bibinfo{person}{Minjune Kim}, \bibinfo{person}{Jin{-}Hee Cho}, \bibinfo{person}{Hyuk Lim}, \bibinfo{person}{Terrence~J. Moore}, \bibinfo{person}{Frederica Free{-}Nelson}, \bibinfo{person}{Ryan K.~L. Ko}, {and} \bibinfo{person}{Dan~Dongseong Kim}.} \bibinfo{year}{2022}\natexlab{}.
\newblock \showarticletitle{Evaluating Performance and Security of a Hybrid Moving Target Defense in {SDN} Environments}. In \bibinfo{booktitle}{\emph{22nd {IEEE} International Conference on Software Quality, Reliability and Security, {QRS} 2022, Guangzhou, China, December 5-9, 2022}}. \bibinfo{publisher}{{IEEE}}, \bibinfo{pages}{276--286}.
\newblock


\bibitem[Lewis et~al\mbox{.}(2020)]%
        {lewis2020retrieval}
\bibfield{author}{\bibinfo{person}{Patrick Lewis}, \bibinfo{person}{Ethan Perez}, \bibinfo{person}{Aleksandra Piktus}, \bibinfo{person}{Fabio Petroni}, \bibinfo{person}{Vladimir Karpukhin}, \bibinfo{person}{Naman Goyal}, \bibinfo{person}{Heinrich K{\"u}ttler}, \bibinfo{person}{Mike Lewis}, \bibinfo{person}{Wen-tau Yih}, \bibinfo{person}{Tim Rockt{\"a}schel}, {et~al\mbox{.}}} \bibinfo{year}{2020}\natexlab{}.
\newblock \showarticletitle{Retrieval-augmented generation for knowledge-intensive nlp tasks}.
\newblock \bibinfo{journal}{\emph{Advances in Neural Information Processing Systems}}  \bibinfo{volume}{33} (\bibinfo{year}{2020}), \bibinfo{pages}{9459--9474}.
\newblock


\bibitem[Lin et~al\mbox{.}(2014)]%
        {Lin_KDD2014}
\bibfield{author}{\bibinfo{person}{Derek Lin}, \bibinfo{person}{Rashmi Raghu}, \bibinfo{person}{Vivek Ramamurthy}, \bibinfo{person}{Jin Yu}, \bibinfo{person}{Regunathan Radhakrishnan}, {and} \bibinfo{person}{Joseph Fernandez}.} \bibinfo{year}{2014}\natexlab{}.
\newblock \showarticletitle{{Unveiling clusters of events for alert and incident management in large-scale enterprise it}}.
\newblock \bibinfo{journal}{\emph{Proceedings of the ACM SIGKDD International Conference on Knowledge Discovery and Data Mining}} (\bibinfo{year}{2014}).
\newblock
\showISBNx{9781450329569}


\bibitem[Liu et~al\mbox{.}(2024)]%
        {Liu_CyberBench2024}
\bibfield{author}{\bibinfo{person}{Zefang Liu}, \bibinfo{person}{Jialei Shi}, {and} \bibinfo{person}{John~F Buford}.} \bibinfo{year}{2024}\natexlab{}.
\newblock \bibinfo{title}{Cyberbench: A multi-task benchmark for evaluating large language models in cybersecurity}.
\newblock
\newblock


\bibitem[Mirheidari et~al\mbox{.}(2013)]%
        {Mirheidari_Note2013}
\bibfield{author}{\bibinfo{person}{Seyed~Ali Mirheidari}, \bibinfo{person}{Sajjad Arshad}, {and} \bibinfo{person}{Rasool Jalili}.} \bibinfo{year}{2013}\natexlab{}.
\newblock \showarticletitle{{Alert correlation algorithms: A survey and taxonomy}}.
\newblock \bibinfo{journal}{\emph{Lecture Notes in Computer Science (including subseries Lecture Notes in Artificial Intelligence and Lecture Notes in Bioinformatics)}}  \bibinfo{volume}{8300 LNCS} (\bibinfo{year}{2013}), \bibinfo{pages}{183--197}.
\newblock
\showISBNx{9783319035833}
\showISSN{03029743}


\bibitem[Senarak(2024)]%
        {senarak2024port}
\bibfield{author}{\bibinfo{person}{Chalermpong Senarak}.} \bibinfo{year}{2024}\natexlab{}.
\newblock \showarticletitle{Port cyberattacks from 2011 to 2023: a literature review and discussion of selected cases}.
\newblock \bibinfo{journal}{\emph{Maritime Economics \& Logistics}} \bibinfo{volume}{26}, \bibinfo{number}{1} (\bibinfo{year}{2024}), \bibinfo{pages}{105--130}.
\newblock


\bibitem[Shittu et~al\mbox{.}(2014)]%
        {SHITTU_LCN2014}
\bibfield{author}{\bibinfo{person}{Riyanat Shittu}, \bibinfo{person}{Alex Healing}, \bibinfo{person}{Robert Ghanea-Hercock}, \bibinfo{person}{Robin Bloomfield}, {and} \bibinfo{person}{Rajarajan Muttukrishnan}.} \bibinfo{year}{2014}\natexlab{}.
\newblock \showarticletitle{{OutMet: A new metric for prioritising intrusion alerts using correlation and outlier analysis}}. In \bibinfo{booktitle}{\emph{Proceedings - Conference on Local Computer Networks, LCN}}. \bibinfo{publisher}{IEEE}, \bibinfo{pages}{322--330}.
\newblock
\showISBNx{9781479937806}


\bibitem[Tihanyi et~al\mbox{.}(2024)]%
        {Tihanyi_CyberMetric2024}
\bibfield{author}{\bibinfo{person}{Norbert Tihanyi}, \bibinfo{person}{Mohamed~Amine Ferrag}, \bibinfo{person}{Ridhi Jain}, {and} \bibinfo{person}{Merouane Debbah}.} \bibinfo{year}{2024}\natexlab{}.
\newblock \showarticletitle{Cybermetric: A benchmark dataset for evaluating large language models knowledge in cybersecurity}.
\newblock \bibinfo{journal}{\emph{arXiv preprint arXiv:2402.07688}} (\bibinfo{year}{2024}).
\newblock


\bibitem[Vaarandi(2021)]%
        {VAARANDI_CSR2021}
\bibfield{author}{\bibinfo{person}{Risto Vaarandi}.} \bibinfo{year}{2021}\natexlab{}.
\newblock \showarticletitle{A stream clustering algorithm for classifying network ids alerts}. In \bibinfo{booktitle}{\emph{2021 IEEE International Conference on Cyber Security and Resilience (CSR)}}. IEEE, \bibinfo{pages}{14--19}.
\newblock


\bibitem[Vaarandi and Podi\v{n}\v{s}(2010)]%
        {VAARANDI_CNSM2010}
\bibfield{author}{\bibinfo{person}{Risto Vaarandi} {and} \bibinfo{person}{K\=arlis Podi\v{n}\v{s}}.} \bibinfo{year}{2010}\natexlab{}.
\newblock \showarticletitle{Network IDS alert classification with frequent itemset mining and data clustering}. In \bibinfo{booktitle}{\emph{2010 International Conference on Network and Service Management}}. \bibinfo{pages}{451--456}.
\newblock


\bibitem[Webb et~al\mbox{.}(2024)]%
        {webb2024cyber}
\bibfield{author}{\bibinfo{person}{Braden~K Webb}, \bibinfo{person}{Sumit Purohit}, {and} \bibinfo{person}{Rounak Meyur}.} \bibinfo{year}{2024}\natexlab{}.
\newblock \showarticletitle{Cyber Knowledge Completion Using Large Language Models}.
\newblock \bibinfo{journal}{\emph{arXiv preprint arXiv:2409.16176}} (\bibinfo{year}{2024}).
\newblock


\end{thebibliography}

\end{document}